\documentclass[conference]{IEEEtran}
\IEEEoverridecommandlockouts
\usepackage{cite}
\usepackage{amsmath,amssymb,amsfonts}
\usepackage{algorithmic}
\usepackage{graphicx}
\usepackage{textcomp}
\usepackage{amsmath}
\usepackage{booktabs}
\usepackage{pifont}
\usepackage{amssymb}
\usepackage{xcolor}
\usepackage{enumitem}
\usepackage{multicol}
\usepackage{multirow}
\usepackage{soul}
\def\BibTeX{{\rm B\kern-.05em{\sc i\kern-.025em b}\kern-.08em
    T\kern-.1667em\lower.7ex\hbox{E}\kern-.125emX}}
\begin{document}
\title{Resilience Revisited: A Multidimensional Framework Derived from Realistic Attack Scenarios\\

\thanks{This work is funded by the National Science Foundation (NSF) Award Number \#2501975.}
}

\author{\IEEEauthorblockN{Isaac Ortega Romero and  Ioannis Zografopoulos}
\IEEEauthorblockA{\textit{Engineering Department, University of Massachusetts Boston, Boston, MA, USA} \\
\textit{email: \{I.OrtegaRomero001 and I.Zografopoulos\}@umb.edu}
}
\vspace{-0.6cm}
}

\maketitle

\begin{abstract}

The increasing penetration of inverter-based resources exposes power systems to coordinated cyber-physical attacks capable of triggering cascading failures and systemic instability. However, existing resilience indicators assess each dimension independently, preventing the quantification of how interdimensional coupling amplifies resilience loss during high-impact, low-probability events. This paper presents a Multidimensional Resilience Index ($\mathcal{MDRI}$) that quantifies degradation across the physical, operational, cyber-digital, climatic, and regulatory dimensions, explicitly distinguishing the independent contribution of each dimension from the additional loss caused by their interactions. The proposed framework is validated on the IEEE 39-bus system implemented in MATLAB/Simulink through two attack scenarios reconstructed from the December 2025 cyberattack on the Polish power grid: a baseline scenario with a single compromised power plant and a coordinated multivector attack. The latter causes system collapse and increases resilience loss nearly eightfold relative to the baseline solely through interdimensional coupling. Including climatic and regulatory stressors produces a further 84\% increase, yielding an overall resilience loss nearly fifteen times greater. These findings demonstrate that multidimensional coupling is a dominant driver of resilience degradation and that resilience assessment must explicitly account for interdependencies among dimensions to reveal vulnerabilities overlooked by conventional approaches.


\end{abstract}


\vspace{-1mm}
\section{Introduction} \label{s:Intro}

The increasing frequency and severity of high-impact, low-probability (HILP) events, including extreme weather, coordinated cyberattacks, and cascading failures, have elevated power system resilience as a critical research priority~\cite{Wang2022A, Yoo2024Modeling}. Unlike reliability, which addresses frequent and predictable contingencies, resilience focuses on the system's ability to limit the extent, systemic impact, and duration of degradations under disturbances~\cite{Wang2022A}. 
This distinction has become increasingly important as modern grids transition toward inverter-based resources, which, despite their role in decarbonization, introduce new cyber-physical attack surfaces that enable coordinated attacks to compromise multiple plants and trigger system-wide instability \cite{ospina2020trustworthy}. The significance of this threat is underscored by a recent joint advisory from the FBI, CISA, NSA, and DOE warning of ongoing state actor-affiliated exploitation of programmable logic controllers across U.S. critical infrastructure sectors~\cite{CISA2026_AA26097A}. These emerging risks highlight the need for assessment frameworks capable of capturing degradation across multiple system dimensions simultaneously~\cite{Sun2024Scenario}.

Prior research has addressed this need through geometric representations such as the resilience triangle and trapezoid, which model degradation and recovery as functions of time. \cite{Mujjuni2021Resilience, Talukder2021Resilience, Yao2022Power}. Furthermore, review studies have classified resilience indicators across technical, social, economic, environmental, and organizational dimensions \cite{zografopoulos2025event, Iswaran2022Power}, 
highlighting the need for comprehensive assessments that span multiple system domains. Building on these foundations, subsequent efforts have expanded the temporal perspective by introducing multi-stage evaluation frameworks in which component criticality evolves dynamically across \emph{i)}~pre-event, \emph{ii)}~during-event, and \emph{iii)}~recovery phases \cite{Stankovic2023Methods,
zografopoulos2022time}. Additionally, probabilistic formulations have been proposed to improve analytical tractability, deriving closed-form resilience metrics under assumptions such as independent failures and additive degradation \cite{Yoo2024Modeling}.

Despite these advances, several fundamental limitations persist. First, geometric resilience curves assume linear performance evolution and can deviate from observed system behavior by up to 22\% \cite{Sun2024Scenario}. Second, most existing metrics focus on a single system dimension, i.e., load curtailment, energy not supplied, network topology, etc., without capturing the multifaceted nature of HILP events \cite{Iswaran2022Power}, and single-dimension assessments have been shown to yield incomplete vulnerability characterizations. Third, existing approaches predominantly treat resilience dimensions as independent, thereby failing to capture cross-dimensional coupling effects. Studies have demonstrated that neglecting interdependencies between system domains can underestimate degradation metrics by as much as 65\% \cite{10221865}, while infrastructure interdependency analyses confirm that HILP events disproportionately amplify impact when multiple dimensions are simultaneously compromised \cite{Kelly-Gorham2024Ranking}. In Table \ref{tab:comparison}, we summarize how related literature addresses these key aspects of resilience assessment.

To address the aforementioned gaps, this paper makes the following contributions:
\begin{itemize}[leftmargin=*]
    \item A multidimensional resilience index ($\mathcal{MDRI}$) that quantifies power system degradation across five dimensions: physical, operational, digital-cyber, climatic, and regulatory within a single comparable metric, separating each dimension's independent contribution from its coupled effect.
    \item A validation study conducted on a 39-bus IEEE system, using two attack scenarios devised from the December 2025 incident on the Polish power grid, which demonstrates the impact of cross-domain coupling effects under coordinated multivector attacks.
\end{itemize}

The remainder of this paper is organized as follows. Section~\ref{s:method} presents the proposed multidimensional resilience framework, Section~\ref{s:results} describes the case study, attack scenarios, and simulation results, and Section~\ref{s:conclusions} concludes the paper. \looseness=-1

\begin{table}[t]
\centering
\caption{Comparison of Resilience Assessment Approaches} \vspace{-4pt}
\label{tab:comparison}
\renewcommand{\arraystretch}{1.15}
\setlength{\tabcolsep}{3pt}
\footnotesize
\begin{tabular}{|c|c|c|c|c|c|}
\hline
\multirow{2}{*}{\textbf{Reference} }& \textbf{Multi-dim.} & \textbf{Number of} & \textbf{Cross-dim.} & \textbf{Exog.} & \textbf{Case} \\
& \textbf{Metrics} & \textbf{Dimensions} & \textbf{Coupling} & \textbf{Factors} & \textbf{Study}\\
\hline
\cite{Mujjuni2021Resilience} & \checkmark & 5 & \ding{55} & \ding{55} & \ding{55} \\
\hline
\cite{Wang2024A} & \checkmark & 2 & \ding{55} & \ding{55} & \checkmark \\
\hline
\cite{Stankovic2023Methods} & \ding{55} & 1 & \ding{55} & \ding{55} & \ding{55} \\
\hline
\cite{Ti2022Resilience} & \checkmark & 2 & \checkmark & \ding{55} & \checkmark \\
\hline
\cite{Dobson2023Models} & \checkmark & 3 & \checkmark & \ding{55} & \checkmark \\
\hline
\cite{Kelly-Gorham2024Ranking}  & \checkmark & 6 & \ding{55} & \ding{55} & \checkmark \\
\hline
\cite{Chivunga2023Power} & \ding{55} & 1 & \ding{55} & \ding{55} & \ding{55} \\
\hline \hline
\textbf{This work} & \checkmark & \textbf{5} & \checkmark & \checkmark & \checkmark \\
\hline
\end{tabular}
\vspace{-5mm}
\end{table}
\vspace{-1mm}

\section{Multidimensional Resilience Framework} \label{s:method}
This section introduces a multidimensional framework for quantifying power system resilience under disturbances. As illustrated in Fig.~\ref{fig:Flowchart}, the proposed approach decomposes system degradation into five dimensions and combines them into a comprehensive resilience index, i.e., $\mathcal{MDRI}$.

\subsection{System Performance and Resilience Loss}

The system performance function $\Phi(t)$, combines frequency deviation and inter-machine coherency, as expressed in Eq.~\eqref{eq:1}:
\vspace{-20pt}

\begin{equation}
\Phi(t)=w_f\Phi_f(t)+w_s\Phi_s(t)
\label{eq:1}
\end{equation}

\vspace{-10pt} 

\begin{equation*}
\begin{aligned}
\Phi_f(t) &= \max \left( 0,\, 1 - \frac{|f_{\mathrm{COI}}(t) - f_{\mathrm{nom}}|}{|f_{\mathrm{nom}} - f_{\mathrm{crit}}|}\right) \\[1pt]
\Phi_s(t) &= \max \left( 0, 1 - \frac{\Delta f_{\mathrm{gen}}(t)}{\Delta f_{\mathrm{gen}}^{\mathrm{coh}}}\right)
\end{aligned}
\vspace{-1pt} 
\end{equation*}

\noindent where $f_{\mathrm{COI}}(t)$ is the center-of-inertia frequency, 
$f_{\mathrm{nom}}$ and $f_{\mathrm{crit}}$ are the nominal and critical frequencies, 
$\Delta f_{\mathrm{gen}}(t)=\max_{i,j}|f_i(t)-f_j(t)|$ is the inter-generator frequency spread, and $\Delta f_{\mathrm{gen}}^{\mathrm{coh}}$ is the coherency tolerance band. The weights satisfy $w_f,w_s\in[0,1]$ and $w_f+w_s=1$. Equal weights are adopted since frequency deviation and coherence loss are considered equally important stability phenomena. These two quantities are selected because they represent the primary indicators of power system stability. Loss of frequency stability or inter-machine coherency typically precedes system collapse, making them the most direct measures of the physical impact of a cyberattack on the power system.

The resilience loss metric $R_{\mathrm{loss}}$ quantifies cumulative performance degradation over a horizon $T_H$ relative to the pre-disturbance operating point. Integration starts at the disturbance time $t_{0,i}$ for scenario $S_i$. To account for collapse, the truncated performance function $\tilde{\Phi}_i(t)$ is defined as:

\begin{equation*}
\tilde{\Phi}_i(t) = 
\begin{cases} 
\Phi_i(t), & t \leq t_{\mathrm{lim}} \\ 
0, & t > t_{\mathrm{lim}}
\end{cases}
\end{equation*}

\noindent where $t_{\mathrm{lim}}=t_f$ if the system recovers within $T_H$, and $t_{\mathrm{lim}}=t_{\mathrm{col}}$ if collapse occurs at $t_{\mathrm{col}}$. The resilience loss is then computed as:

\begin{figure}[t!] \vspace{-1mm}
\centerline{\includegraphics[trim=0 20pt 0 0pt, clip, width=1\columnwidth]{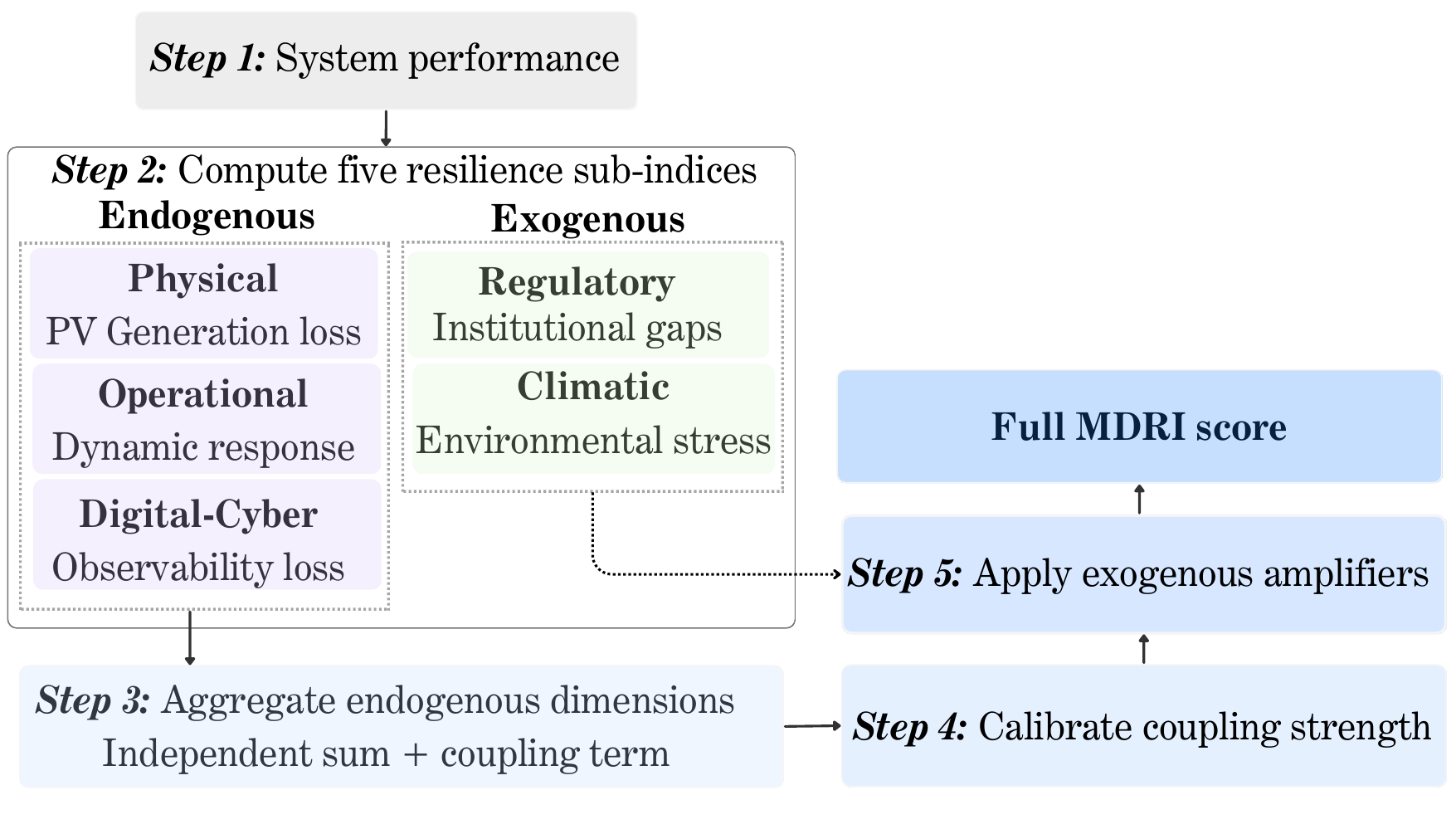}}
\caption{Algorithmic overview of the proposed MDRI score calculation.}
\label{fig:Flowchart}
\vspace{-4mm}
\end{figure}

\vspace{-15pt}

\begin{equation}
R_{\mathrm{loss},i} = \frac{1}{\Phi_{0,i}} \int_{t_{0,i}}^{t_{0,i}+T_H} 
\max\!\left(0,\; \Phi_{0,i} - \tilde{\Phi}_i(t)\right) dt
\label{eq:2}
\end{equation}

\noindent where $\Phi_{0,i}$ denotes the pre-disturbance performance level for scenario $S_i$.

\subsection{Resilience Dimension Definitions}

\subsubsection{Physical Dimension}
The physical disruption index $D_{\mathrm{phy},i}$ in Eq.~\eqref{eq:3} quantifies the weighted loss of generation capacity caused by a disturbance in scenario $S_i$:
\vspace{-5pt}

\begin{equation}
D_{\mathrm{phy},i} = \sum_{r \in R} \omega_r \frac{P_{\mathrm{r,lost},i}}
{P_{\mathrm{r,total}}}
\label{eq:3}
\end{equation}

\noindent where $P_{\mathrm{r,lost},i}$ is the disconnected or unavailable capacity of resource $r$ in scenario $S_i$ (MW), $P_{\mathrm{r,total}}$ is its installed capacity (MW), $\omega_r\in[0,1]$ is the assigned weight, and $R=\{\mathrm{PV,\, synchronous,\, storage,\, substations,\dots}\}$ denotes the set of resource types. The weights reflect the relative importance of affected resources in $S_i$, while unaffected types are excluded. When multiple resources are impacted, weights are assigned according to their criticality (i.e., inertia, reserves, or critical-load support). If only one resource drives the disruption, $\omega_r=1$.

\subsubsection{Operational Dimension}
The operational disruption index $D_{\mathrm{op},i}$ characterizes the system dynamic response to a disturbance in scenario $S_i$ by combining frequency variation rate, performance degradation, and generator coherency, as defined in Eq.~\eqref{eq:4}:
\vspace{-10pt}

\begin{equation}
\begin{gathered}
D_{\mathrm{op},i} = \frac{X_i}{1+X_i}, \\[.8ex]
X_i = \frac{1}{3}\left(
\frac{|\mathrm{RoCoF}|_{\max,i}}{\mathrm{RoCoF}_{\mathrm{crit}}}
+\frac{\delta_{\Phi,i}}{\delta_{\Phi}^{\mathrm{crit}}}
+\frac{\Delta f_{\mathrm{gen},i}^{\max}}
{\Delta f_{\mathrm{gen}}^{\mathrm{crit}}}
\right)
\end{gathered}
\label{eq:4}
\end{equation}

\begin{equation*}
\delta_{\Phi,i} =
\frac{\Phi_{0,i} - \Phi_{\text{nadir},i}}{\Phi_{0,i}}
\end{equation*}

\noindent where $|\mathrm{RoCoF}|_{\max,i}$ is the maximum absolute rate of change of frequency, $\delta_{\Phi,i}$ is the normalized performance drop, $\Phi_{\mathrm{nadir},i}$ is the minimum value of $\Phi(t)$ during the event, and $\Delta f_{\mathrm{gen},i}^{\max}=\max_t \Delta f_{\mathrm{gen},i}(t)$ is the peak inter-generator frequency spread. Equal weights are assigned to the three terms due to their complementary role in transient stability degradation. Critical thresholds are set to $\mathrm{RoCoF}^{\mathrm{crit}}=1.0$~Hz/s \cite{entsoe2023inertia}, $\delta_{\Phi}^{\mathrm{crit}}=0.05$, and $\Delta f_{\mathrm{gen}}^{\mathrm{crit}}=2.0$~Hz.

The saturating form bounds $D_{\text{op},i}$ within $[0,1)$ while preserving sensitivity near the critical region ($X_i=1 \to D_{\text{op},i}=0.5$). As $X_i$ increases, $D_{\text{op},i}$ asymptotically approaches unity.

\subsubsection{Digital-Cyber Dimension}
The digital-cyber disruption index $D_{\mathrm{cyb},i}$ quantifies the impact of a disturbance on system during scenario $S_i$, incorporating observability, controllability, integrity, and availability into the normalized index of Eq.~\eqref{eq:5}:
\vspace{-15pt}

\begin{equation}
D_{\mathrm{cyb},i} = \sum_{j \in \mathcal{K}} w_{\mathrm{cy}_j} 
\cdot \frac{N_{\mathrm{compr}_j,i}}{N_{\mathrm{scope}_j}}
\label{eq:5}
\end{equation}

\noindent where $\mathcal{K}=\{\mathrm{obs,\, ctrl,\, int,\, av}\}$ denotes the evaluated cyber aspects, $N_{\mathrm{compr}_j,i}$ is the number of compromised assets in aspect $j$ for scenario $S_i$, $N_{\mathrm{scope}_j}$ is the total number of evaluated assets, and $w_{\mathrm{cy}_j}\geq0$ are weighting factors satisfying $\sum_{j \in \mathcal{K}} w_{\mathrm{cy}_j}=1$.

\subsubsection{Climatic Dimension}

The climatic disruption index $D_{\mathrm{clim},i}$ represents environmental stressors that exacerbate system degradation during scenario $S_i$. Multiple climatic factors are aggregated into the normalized index of Eq.~\eqref{eq:6}:
\vspace{-5pt}

\begin{equation}
D_{\mathrm{clim},i} = \sum_{k \in \mathcal{C}} w_{c_k} \cdot I_{c_k,i}
\label{eq:6}
\end{equation}

\noindent where $C=$\{temperature, snow, wind, ice, humidity, extreme weather, \dots\} is the set of climatic stressors considered, $I_{c_k,i}\in[0,1]$ is the normalized intensity of stressor $k$ in scenario $S_i$, and $w_{c_k}\geq0$ are weighting factors satisfying $\sum_{k \in \mathcal{C}} w_{c_k}=1$.

\subsubsection{Regulatory Dimension} The regulatory dimension quantifies institutional vulnerabilities by comparing the existing controls against a reference framework. The sub-index is expressed in Eq.~\eqref{eq:7} as:

\begin{equation}\vspace{-1pt}
    D_{\text{reg}} = \frac{1}{N_v^{\text{ref}}} 
    \sum_{i=1}^{N_v^{\text{ref}}} v_i
    \label{eq:7}
    \vspace{-5pt}
\end{equation}

\noindent where $v_i \in \{0,1\}$ indicates the presence ($v_i = 1$) 
or absence ($v_i = 0$) of a regulatory weakness in the control category $i$, and $N_v^{\text{ref}}$ is the total number of reference control categories, selected independently of the case under study.

\subsection{Multidimensional Resilience Index}
The proposed formulation is based on the premise that evaluating resilience dimensions independently underestimates systemwide impacts, i.e.,  their simultaneous compromise and underlying interdependencies create degradation that no \textit{single-dimensional} assessment can capture.

Let $\mathcal{K}_{\mathrm{sim}} = \{\mathrm{phy,\, op,\, cyb}\}$ 
be the set of endogenous dimensions. The endogenous core decomposes degradation into an \textbf{additive} and a \textbf{coupling} contribution, as given by eq.~\eqref{eq:10}:
\vspace{-8pt}
\begin{equation}
\mathcal{M}(S_i;\gamma_i) =
\underbrace{%
  \frac{1}{|\mathcal{K}_{\mathrm{sim}}|}
  \sum_{k \in \mathcal{K}_{\mathrm{sim}}} D_{k,i}
}_{\overline{D}_i \; (\text{additive})}
+\;
\gamma_i
\underbrace{%
  \prod_{k \in \mathcal{K}_{\mathrm{sim}}} D_{k,i}
}_{\Pi_i \; (\text{coupling})}
\label{eq:10}
\end{equation}

\noindent where $D_{k,i} \in [0,1]$ is the normalized sub-index of dimension $k$ in scenario $S_i$, with equal weights $1/|\mathcal{K}_{\mathrm{sim}}|$. The additive term 
$\bar{D}_i$ measures mean severity across dimensions independently. The coupling term $\Pi_i = \prod_{k} D_{k,i}$ captures additional degradation arising from simultaneous 
cross-dimensional compromise; it collapses to zero whenever any single dimension remains uncompromised ($D_{k,i} \to 0$), and reaches its maximum only when all dimensions are 
jointly and severely degraded, encoding the cascading failure mechanism whereby an intact dimension suppresses impact propagation, while simultaneous degradation across 
all dimensions produces mutual amplification beyond the additive prediction.

The parameter $\gamma_i$ takes one of two values. Scenarios driven by a single disturbance vector are assigned 
to the \textbf{additive regime} ($\gamma_i = 0$), under which $\mathcal{M}(S_i;0) = \bar{D}_i$ and no cross-dimensional interaction is considered. Scenarios where all 
endogenous dimensions are simultaneously compromised are assigned to the \textbf{coupled regime} ($\gamma_i = 1$), activating $\Pi_i$ and amplifying degradation beyond the additive baseline.
The $\mathcal{MDRI}$ for scenario $S_i$ is given by Eq.~\eqref{eq:11}:
 \vspace{-2pt}
\begin{equation}
\mathcal{MDRI}_i = \mathcal{M}(S_i;\gamma_i) \cdot
\prod_{j \in \mathcal{K}_{\mathrm{ext}}} (1 + D_{j,i})
\label{eq:11}
\end{equation}
 
\noindent where $\mathcal{K}_{\mathrm{ext}} = \{\mathrm{clim,\, reg}\}$ is the set of exogenous dimensions and $D_{j,i} \geq 0$ is the normalized sub-index of exogenous dimension $j$ in scenario $S_i$. Each factor $(1 + D_{j,i})$ amplifies the endogenous core proportionally to the exogenous stress level. When no exogenous stress is present ($D_{j,i} = 0$), the factor reduces to unity, and the index simplifies to $\mathcal{MDRI}_i = \mathcal{M}(S_i;\gamma_i)$. Since the exogenous amplifiers act multiplicatively on the endogenous core, $\mathcal{MDRI}_i$ is not normalized to $[0,1]$ and serves as a comparative metric across scenarios.

\section{Results} \label{s:results}

\subsection{Case Study and Attack Scenarios} \label{s:system_models}
\noindent\textbf{System Model:~}The proposed framework is validated on the IEEE 39-bus test system, implemented in MATLAB/Simulink. The original system comprises 10 synchronous generators with a total installed capacity of 10,610 MW. To represent the increasing penetration of inverter-based resources, 1,500 MW of synchronous generation has been replaced by nine utility-scale grid-forming PV plants distributed across buses $30-38$ (14.1\% of total capacity). 

\noindent\textbf{Threat Model:~}Assumes a state-sponsored adversary targeting both the information technology (IT) and operational technology (OT) domains, consistent with the tactics, techniques, and procedures (TTPs) attributed to the ELECTRUM/Sandworm group~\cite{CERTPolska2026, Dragos2026, ESET2026}. The attacker possesses prior knowledge of grid topology and industrial control system (ICS) protocols, exploiting exposed perimeter devices and default credentials for initial access and lateral movement into OT networks. Attack execution involves coordinated, multi-vector actions across distributed sites, including control signal manipulation and communication disruption.
 
\noindent\textbf{Attack Scenarios:~}Two distinct attack scenarios are evaluated, inspired by the coordinated cyberattack on the Polish energy infrastructure on December 29, 2025~\cite{CERTPolska2026}. 

\textbf{Scenario A} -- Single-plant baseline attack:~A control input attack manipulates the active power reference of the PV plant at Bus 33 (190~MW), selected for its proximity to the highest-load buses in the system. As a result, a generation loss at this bus produces measurable system-wide frequency transients while remaining within the single-plant scope for our baseline scenario. Operators retain full communication and control over all the remaining plants. This represents an isolated cyberattack without any climatic or regulatory assumptions.

\textbf{Scenario B} -- Multi-vector cascading attack:~A coordinated attack replicates TTPs documented in the Polish grid incident: \emph{i)} communication disruption targeting six PV plants, eliminating operator observability and control, and \emph{ii)} forced disconnection of 1,115~MW of PV capacity (74.3\% of total). The six targeted plants (at Buses 31, 32, 34, 35, 37, and 38) are geographically dispersed, mirroring the targeting strategy. To reflect the elevated winter demand observed during the Polish incident, a $25\%$ load increase is introduced as an operational stress factor, introduced by the sub-zero temperatures and snowstorms~\cite{CERTPolska2026, Dragos2026, zografopoulos2020special}.
\vspace{-5pt}

\subsection{Attack Impact and Dynamic Response}
\begin{figure}[t]
\centerline{\includegraphics[width=0.94\columnwidth]{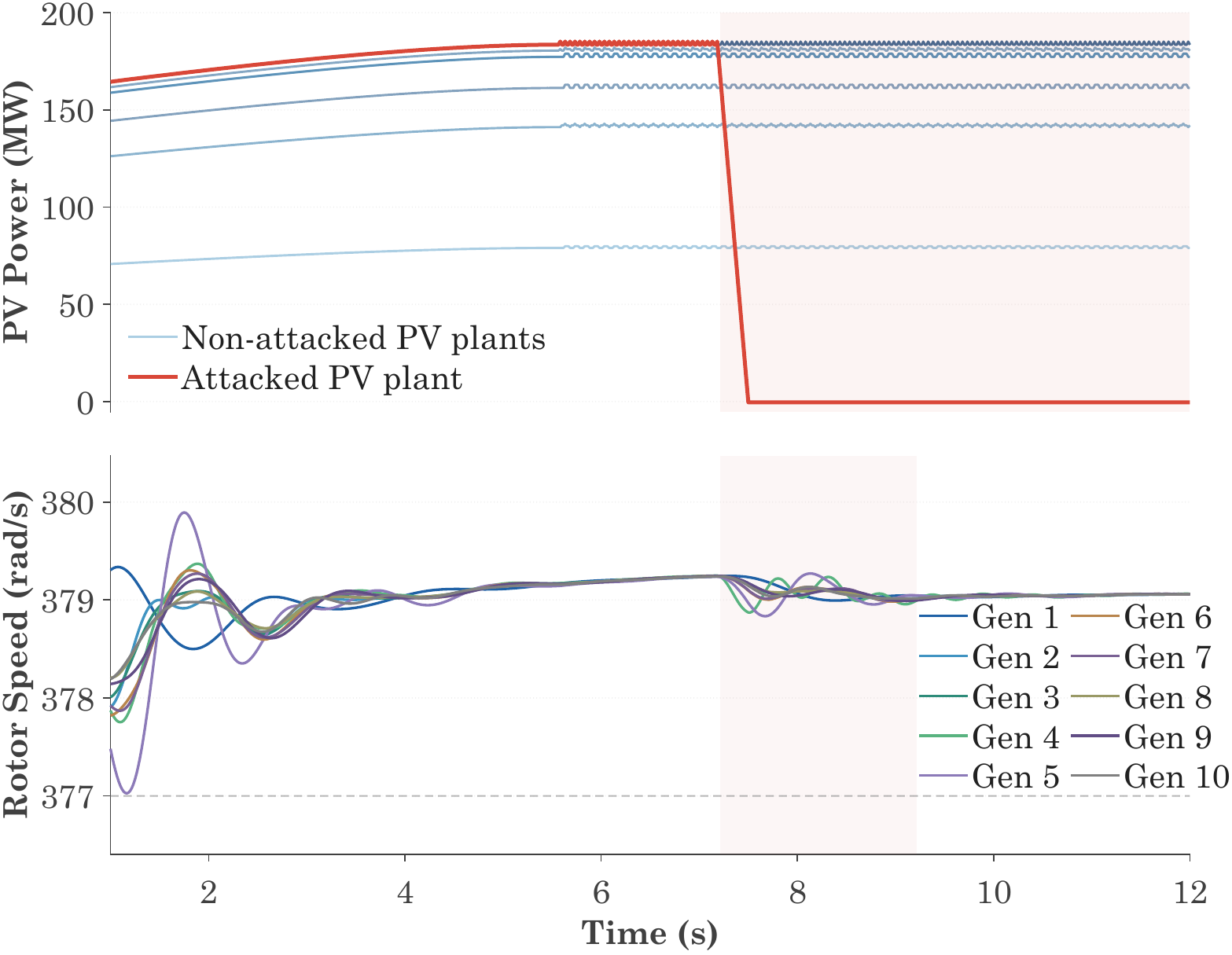}} \vspace{-8pt}
\caption{Scenario A: PV generation and rotor speed response.}
\label{fig:AttackSA}
    \vspace{-10pt}
\end{figure}

\begin{figure}[t]
\centerline{\includegraphics[width=0.92\columnwidth]{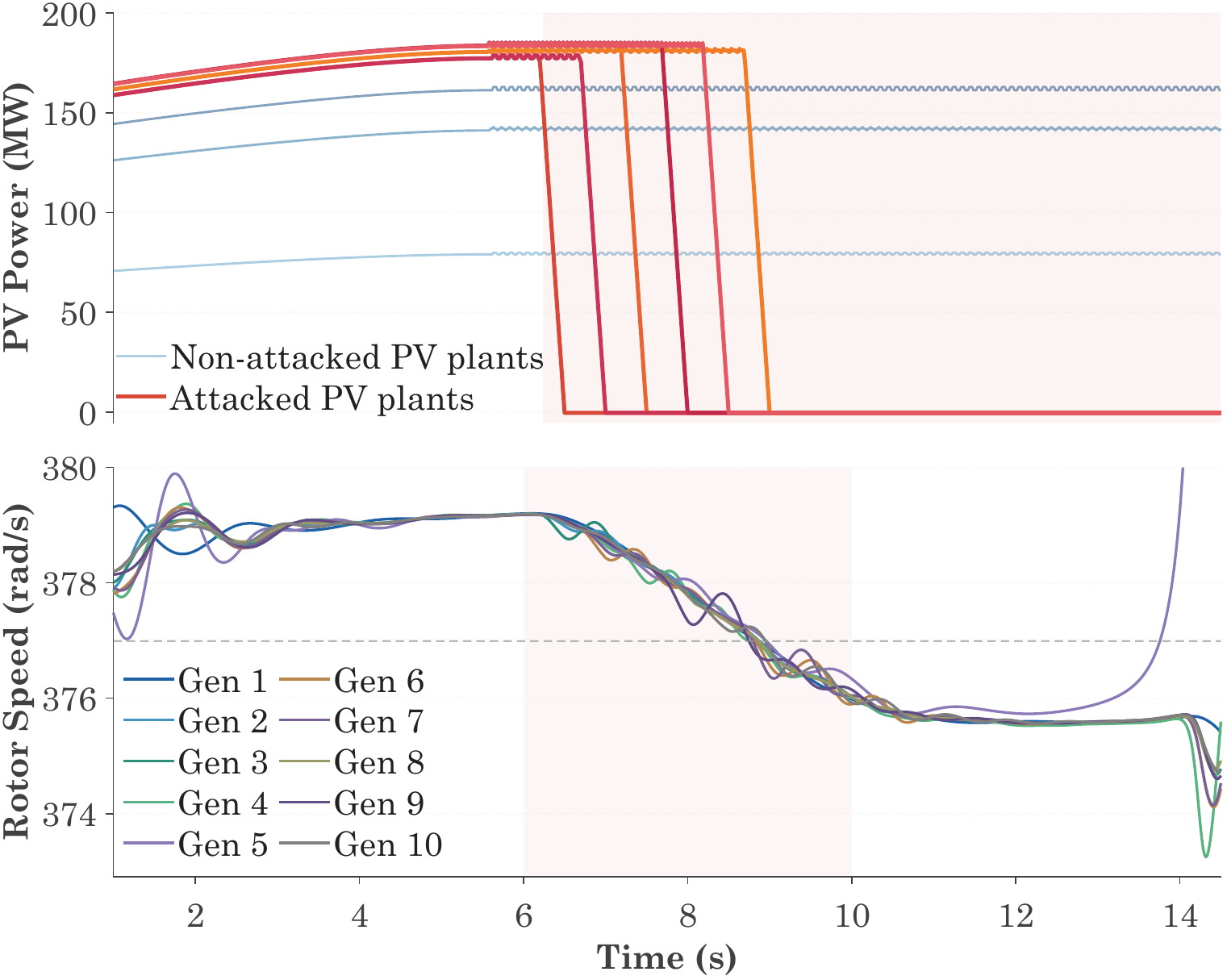}} \vspace{-8pt}
\caption{Scenario B: PV generation and rotor speed response.}
\label{fig:AttackSB}
    \vspace{-8pt}
\end{figure}

Fig.~\ref{fig:AttackSA} and \ref{fig:AttackSB} present the PV generation output and synchronous generator rotor speeds for both scenarios, while Table~\ref{tab:frequency_metrics} summarizes the frequency response for both scenarios.

\noindent \textbf{Scenario A:}~As demonstrated in Fig.~\ref{fig:AttackSA}, the attack on the PV plant forces its output from 190~MW to zero (at $t=7s$), while non-attacked plants maintain nominal generation. All generators exhibit brief transient oscillations before converging to a common steady-state speed.

During Scenario A, the system maintains frequency stability with negligible frequency deviations, indicating full inter-machine coherency throughout the transient. Bus voltages remain within nominal ranges, and no load shedding is triggered.\looseness=-1

\noindent\textbf{Scenario B:}~As demonstrated in Fig.~\ref{fig:AttackSB}, six PV plants are simultaneously disconnected, removing 1,115~MW from the system. The rotor speed responses reveal dynamic instabilities, i.e.,  generators attempt a coordinated response, but beyond $t\approx8s$ their trajectories diverge, leading to loss of synchronism.

During Scenario B, frequency performance deteriorates, with a 61.8\% increase in maximum RoCoF compared to Scenario A. The generator frequency fluctuations reach $3.10$Hz, indicating complete loss of synchronism among synchronous machines. As shown in Fig.~\ref{figure1}, widespread voltage sags are observed and the voltage angle dispersion increases substantially, reflecting the loss of angular coherency across the system.

\begin{table}[t]
\centering
\caption{Frequency response metrics comparison} \vspace{-4pt}
\begin{tabular}{|l|c|c|}
\hline
\textbf{Metric} & \textbf{Scenario A} & \textbf{Scenario B} \\
\hline
Nadir frequency [Hz]        & 60.323    & 59.776  \\
Time to frequency nadir [s]           & 8.72      & 12.82   \\
Maximum RoCoF [Hz/s]        & -0.055    & -0.089  \\
RoCoF time [s]              & 7.5       & 6.5     \\
Steady-state frequency [Hz] & 60.332    & \textit{unstable} \\
$\Delta f_{\mathrm{gen}}$ [Hz]       & 0.00012   & 3.10    \\
\hline
\end{tabular}
\label{tab:frequency_metrics}  \vspace{-14pt}
\end{table}

\begin{figure}[t]
\centerline{\includegraphics[width=0.9\columnwidth]{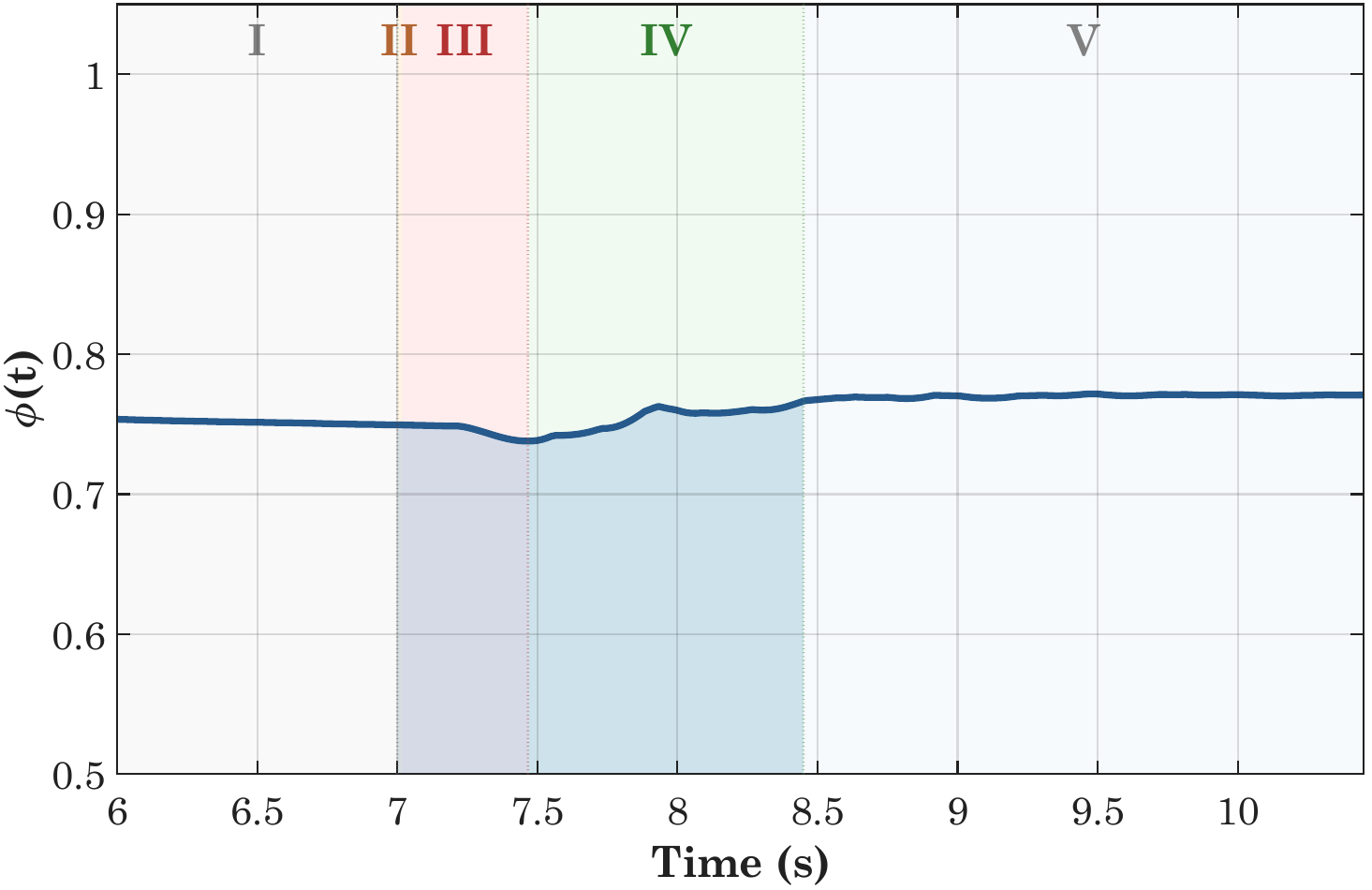}} \vspace{-8pt}
\caption{System performance function and phases for Scenario A.}
\label{fig:resilience_S1}  \vspace{-8pt}
\end{figure}
\vspace{-0.8pt}

\subsection{System Performance and Resilience Curves}

Fig.~\ref{fig:resilience_S1} and \ref{fig:resilience_S2} present \(\Phi(t)\) for scenarios A and B (\(w_f\!=\!w_s\!=\!0.5\), 
\(f_{\mathrm{nom}}\!=\!60\)Hz, \(f_{\mathrm{crit}}\!=\!59\)Hz, \(\Delta f_{\mathrm{gen}}^{\mathrm{coh}}\!=\!1.0\)Hz, \(T_H\!=\!15\)s), where $\Phi_0\!<\!1$ due to residual PV-integration deviations.

\textbf{Scenario A} -- Successful Recovery:
Following the attack on a single PV plant, the system performance degrades from $\Phi_0 = 0.7674$ to a nadir of $0.7379$ at $t = 7.47s$, representing a 3.8\% drop. Recovery to $\Phi = 0.7670$ (99.9\% of pre-event level) occurs within $1.45s$, and $R_{\text{loss}} =0.0305$. The system performance function exhibits five distinct phases: \emph{i)} pre-event steady state, \emph{ii)} absorption ($0.02s$), \emph{iii)} degraded operation ($0.45s$), \emph{iv)} recovery ($0.98s$), and \emph{v)} post-event equilibrium.

\begin{figure}[t]
\centerline{\includegraphics[width=0.9\columnwidth]{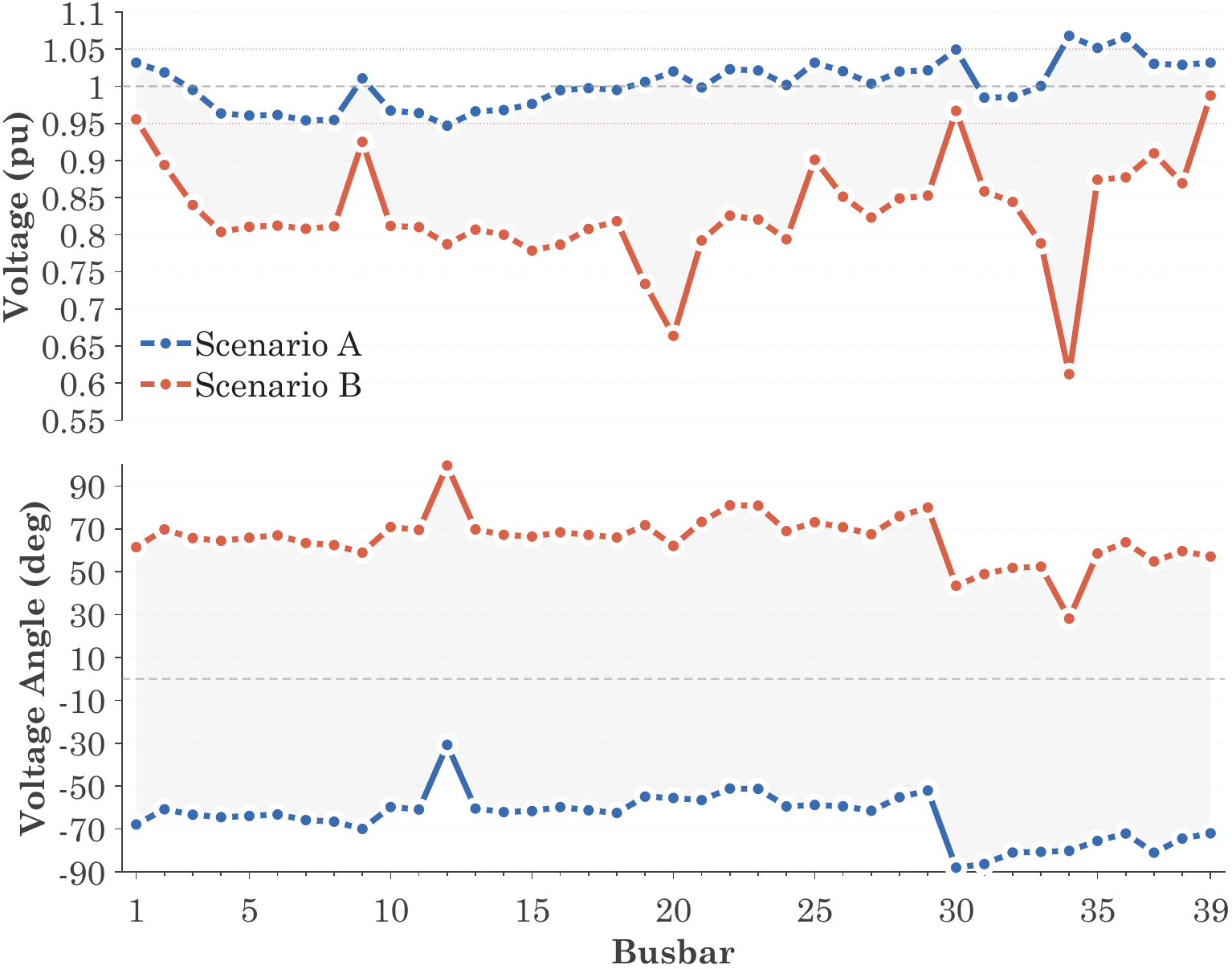}} \vspace{-8pt}
\caption{Voltage magnitudes and angle profiles across all 39 buses at $t=15s$.} \vspace{-10pt}
\label{figure1}
\end{figure}

\textbf{Scenario B} -- Cascading System Collapse:
The multi-vector attack triggers progressive degradation from $\Phi_0 = 0.7674$ to a nadir of $0.5549$ at $t = 14.57s$, a 27.7\% loss. The system does not recover and the collapse is detected at $t = 14.22s$ with $\Phi = 0.5824$, and $R_{\text{loss}} =1.0080$. The system performance exhibits only four phases (absent post-event equilibrium): \emph{i)} pre-event steady state, \emph{ii)} attack propagation ($4.00s$), during which the system attempts to absorb the generation loss over approximately $3s$ before losing stability entirely, \emph{iii)} degraded operation ($4.22s$), \emph{iv)} failed recovery ($0.35s$). The short-lived phase \emph{iv} fails to mitigate the cascading instability, leading to complete system collapse at $t=14.22s$. \looseness=-1

\begin{figure}[t]
\centerline{\includegraphics[width=0.98\columnwidth]{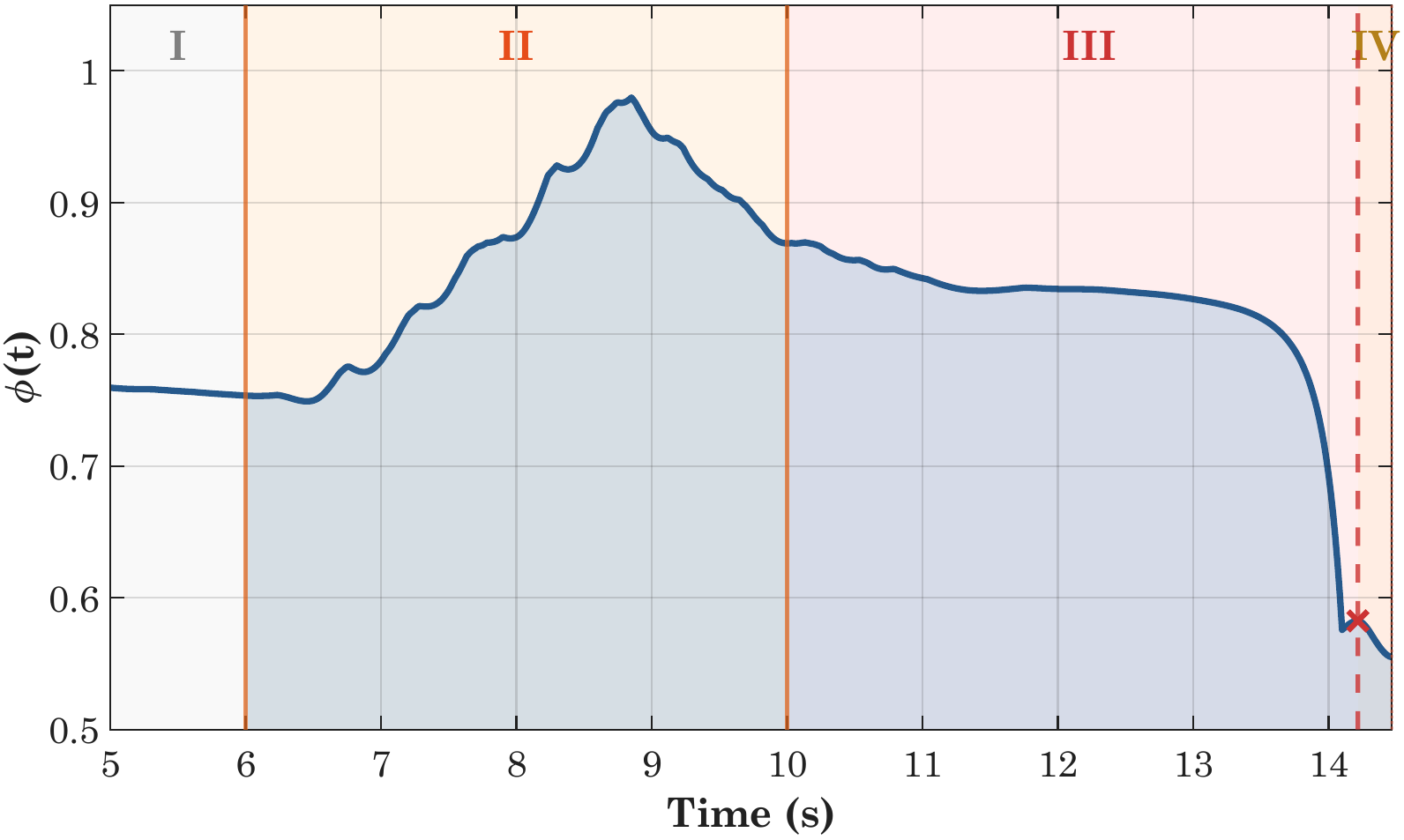}} \vspace{-8pt}
\caption{System performance function and phases for Scenario B.}
\label{fig:resilience_S2}  \vspace{-8pt}
\end{figure}

\subsection{$\mathcal{MDRI}$ Evaluation}
\noindent Table~\ref{tab:subindices} summarizes the physical, operational, and digital-cyber sub-indices together with their corresponding average degradation, $\bar{D}_i$, and interaction product, $\Pi_i$.

\textbf{Physical and Operational Dimensions:} The physical sub-index, $D_{\text{phy},i}$, measures the fraction of inverter-based generation capacity lost. Since all disturbances exclusively affect the PV fleet, Eq.~\eqref{eq:3} reduces to $D_{\text{phy},i}=\frac{P_{\text{PV,lost},i}}{P_{\text{PV,total}}}$ with $\omega_{\text{PV}}=1$. This yields $D_{\text{phy},A}=0.127$ and $D_{\text{phy},B}=0.743$.

The operational sub-index, $D_{\text{op},i}$ in Scenario~A, indicators remain below their critical thresholds, resulting in $D_{\text{op},A}=0.214$. In Scenario~B, the simultaneous disconnection of multiple PV plants causes both $\delta_{\Phi}$ and $\Delta f_{\text{gen}}^{\max}$ to exceed their critical limits. Specifically, $\delta_{\Phi,B}=0.277$ exceeds $\delta_{\Phi}^{\text{crit}}$ by a factor of 5.5, while the inter-generator frequency spread reaches $3.097$~Hz, surpassing the 2.0~Hz coherency threshold, as a result, $D_{\text{op},B}=0.705$.

\textbf{Digital-Cyber Dimension:} The digital-cyber sub-index, $D_{\text{cyb},i}$, is computed from four complementary aspects: observability, controllability, integrity, and availability, each assigned an equal weight of $w_{\text{cy}j}=0.25$. Equal weighting reflects their non-redundant contributions to characterizing operator situational awareness, remote control capability, data trustworthiness, and asset accessibility during cyber incidents. The assessment scope is defined as $N{\text{scope}}=9$, corresponding to the nine PV plants comprising the inverter-based fleet.

Under Scenario~A, the attack manipulates the active power reference of a single PV plant, compromising only integrity and availability ($N_{\text{compr,int}}=N_{\text{compr,av}}=1$), while observability and controllability remain unaffected ($N_{\text{compr,obs}}=N_{\text{compr,ctrl}}=0$). This yields $D_{\text{cyb},A}=0.056$. In contrast, Scenario~B compromises communication, supervisory interfaces, firmware, and plant availability at the six targeted sites ($N_{\text{compr,obs}}=N_{\text{compr,ctrl}}=N_{\text{compr,int}}=N_{\text{compr,av}}=6$), resulting in $D_{\text{cyb},B}=0.667$, a more than tenfold increase that reflects the broader cyber impact of the coordinated attack.

\textbf{Climatic-External Dimension:} Scenario~A assumes nominal weather conditions, yielding $D_{\text{clim}}^A=0$. In Scenario~B, the cyber-physical disturbance coincides with adverse winter conditions representative of the Polish event. The climatic sub-index considers the latter two exogenous stressors, which are normalized and equally weighted ($w_{\text{temp}}=w_{\text{snow}}=0.5$): thermal stress due to sub-zero ambient temperature, normalized using the IEC~60076 thermal reference ($I_{\text{temp}}=0.20$), and additional mechanical loading on non-attacked PV plants caused by snowfall, normalized according to the EN~1991-1-3 snow load standard ($I_{\text{snow}}=0.10$). These values yield $D_{\text{clim}}^B=0.150$.

\textbf{Regulatory~Dimension:} 
Based on the mapping of IEC~62443 and NERC~CIP, a reference set of $N_v^{\text{ref}} = 10$ control categories is defined. The Polish attack compromised six of these~\cite{CERTPolska2026, Dragos2026, Zetter2026,ESET2026}: \textit{i)} no multi-factor authentication on FortiGate virtual private networks, \textit{ii)} unpatched firmware with known exploitable vulnerabilities, \textit{iii)} default/reused credentials on Hitachi remote terminal units and Mikronika controllers, \textit{iv)} poor IT/OT segmentation, \textit{v)} non-compliance with mandatory distributed energy resecources  cybersecurity standards, \textit{vi)} insufficient OT monitoring at remote substations. The remaining four (i.e., incident response, supply chain risk, physical security, awareness) were unaffected. Thus, $D_{\text{reg}} = 0$ for scenario A, and $D_{\text{reg}} = 0.6$ for scenario B. Based on  Table~\ref{tab:subindices}, Scenario~A is classified under the additive ($\gamma_A=0$), whereas Scenario~B operates in the coupled ($\gamma_B=1$) regime.

\begin{table}[t]
\footnotesize
\caption{Endogenous index values per scenario.}
\label{tab:subindices}
\centering
\begin{tabular}{lcc}
\toprule
Quantity & $S_A$ & $S_B$ \\
\midrule
$D_{\mathrm{phy},i}$ & 0.127 & 0.743 \\
$D_{\mathrm{op},i}$  & 0.214 & 0.705 \\
$D_{\mathrm{cyb},i}$ & 0.056 & 0.667 \\
$\bar{D}_i$          & 0.132 & 0.705 \\
$\Pi_i=\prod_k D_{k,i}$ & 0.0015 & 0.349 \\
\midrule
Regime & additive & coupled \\
$\gamma_i$ & 0 & 1 \\
\bottomrule
\end{tabular}
\vspace{-3mm}
\end{table}
\vspace{-1pt}

In Scenario~A, the small cyber degradation ($D_{\mathrm{cyb},A}=0.056$) leads to a negligible interaction product ($\Pi_A=0.0015$), making the endogenous degradation essentially additive. By contrast, the simultaneous increase of all three sub-indices in Scenario~B yields $\Pi_B=0.349$, causing the coupling term to account for approximately $33\%$. This result highlights the importance of cross-dimensional interactions under coordinated disturbances. The resulting $\mathcal{MDRI}$ values, are reported in Table~\ref{tab:4}.

\begin{table}[t]
\footnotesize
\caption{$\mathcal{MDRI}$ values per scenario.}
\label{tab:4}
\centering
\begin{tabular}{lcc}
\toprule
Component & $S_A$ & $S_B$ \\
\midrule
$\mathcal{M}(S_i;\gamma_i)$ & 0.132 & 1.054 \\
$(1+D_{\mathrm{clim},i})$ & --- & 1.150 \\
$(1+D_{\mathrm{reg},i})$ & --- & 1.600 \\
\midrule
$\mathcal{MDRI}_i$ & 0.132 & 1.940 \\
\bottomrule
\end{tabular}
\vspace{-3mm}
\end{table}

The increase in degradation between scenarios can be expressed through a logarithmic decomposition as: 
\vspace{-1pt}
\begin{equation}
    \ln\!\left(\frac{\mathcal{MDRI}_B}{\mathcal{MDRI}_A}\right)
=
\ln\!\left(\frac{\mathcal{M}_B}{\mathcal{M}_A}\right)
+
\sum_{k\in\mathcal{K}_{\mathrm{ext}}}
\ln\!\left(
\frac{1+D_{k,B}}
     {1+D_{k,A}}
\right)
\end{equation}

The decomposition shows that $77\%$ of the increase in $\mathcal{MDRI}$ from Scenario~A to Scenario~B is explained by the endogenous core through cross-dimensional coupling, while the remaining $23\%$ arises from climatic and regulatory effects.

\section{Conclusion} \label{s:conclusions}
This paper proposes a multidimensional resilience framework, $\mathcal{MDRI}$, that decomposes power system degradation into five dimensions to characterize vulnerability under coordinated cyberattacks. By separating each dimension's independent contribution from its coupled effect, the MDRI captures degradation that single-dimensional assessments cannot. Validation on the IEEE 39-bus system under two scenarios inspired by the December 2025 Polish cyberattack demonstrates that a coordinated multi-vector attack raises the endogenous core roughly $8\times$ over a single-vector baseline through cross-dimensional coupling alone, while climatic and regulatory stressors add a further 84\%, yielding an approximately $15\times$ overall increase in resilience loss, confirming that resilience quantification cannot be decoupled from institutional and environmental contexts.



\bibliographystyle{IEEEtran}
\bibliography{references}

\end{document}